\documentclass[aps,pre,reprint,showpacs,bibnotes,twoside,twocolumn,eqsecnum]{revtex4}

\usepackage{amsmath}
\usepackage{amsfonts}
\usepackage{amssymb}
\usepackage{graphicx}
\usepackage{natbib}
\usepackage{color}

\providecommand{\U}[1]{\protect\rule{.1in}{.1in}}

\begin{document}

\title{Iterative Method for Generating Correlated Binary Sequences}

\author{O.~V.~Usatenko}
\email{usatenko@ire.kharkov.ua}
\affiliation{A.~Ya.~Usikov Institute for Radiophysics and Electronics,\\ Ukrainian Academy of Science, 12 Proskura Street, 61085 Kharkov, Ukraine}

\author{S.~S.~Melnik}
\affiliation{A.~Ya.~Usikov Institute for Radiophysics and Electronics,\\ Ukrainian Academy of Science, 12 Proskura Street, 61085 Kharkov, Ukraine}

\author{S.~S.~Apostolov}
\affiliation{A.~Ya.~Usikov Institute for Radiophysics and Electronics,\\ Ukrainian Academy of Science, 12 Proskura Street, 61085 Kharkov, Ukraine}

\author{N.~M.~Makarov}
\affiliation{Instituto de Ciencias, Benem\'erita Universidad Aut\'{o}noma de Puebla,\\ Priv. 17 Norte No. 3417, Col. San Miguel Hueyotlipan, Puebla, Pue., 72050, M\'{e}xico}

\author{A.~A.~Krokhin}
\affiliation{Department of Physics, University of North Texas, P.O. Box 311427, Denton, TX 76203, USA}

\date{\today}

\begin{abstract}
We propose a new efficient iterative method for generating random correlated binary sequences with prescribed correlation function. The method is based on consecutive linear modulations of initially uncorrelated sequence into a correlated one. Each step of modulation increases the correlations until the desired level has been reached. Robustness and efficiency for the proposed algorithm are tested by generating sequences with inverse power-law correlations. The substantial increase in the strength of correlation in the iterative method with respect to the single-step
filtering generation is shown for all studied correlation functions. Our results can be used for design of disordered superlattices, waveguides, and surfaces with selective transport properties.
\end{abstract}


\maketitle

\section{Introduction}
\label{sec-intro}

Generation of random numbers is a serious mathematical and numerical problem. Any numerical algorithm generates a finite-length pseudo-random sequence where correlations are inevitable.  One of the quality factors of a random number generator includes strength of correlations since even quite weak correlations may lead to untruthful results obtained by the Monte-Carlo method~\cite{Comp}. However, random sequences with prescribed correlations are necessary for design of random lasers~\cite{Cao}, waveguides and surfaces with selective transport properties~\cite{Sto,Mar,IzKr99}, and for analysis of anomalous diffusion~\cite{Gri}. An uncorrelated sequence can be converted into a correlated one using one of the known algorithms: the Mandelbrot fast fractional gaussian noise generation~\cite{Mand71}, the Voss procedure of consequent
random addition~\cite{voss}, the correlated Levy walks~\cite{shl}, the convolution method (often referred to as Rice's algorithm or inverse Fourier transformation)~\cite{Rice,Saupe88Feder88,WOD95,IzKr99,czir,RewUAMM}. The latter algorithm may generate a sequence possessing practically \emph{any} correlation function which is allowed by statistics. This universality is a manifestation of the continuity of the space of states -- set of the real numbers $a(n) \in R, \,\,\, (-\infty< a(n)< \infty)$ -- the terms of the generated sequence belong to.

If, however, the terms of a sequence are allowed to take only some discrete values, like it occurs, for example, in a sequence of nucleotides in a DNA molecule, the problem of generating either a truly random sequence or a sequence with prescribed correlations becomes more complicated.
The limiting case of a sequence with discrete space of states is a sequence written by two symbols $0$ and $1$. Properties and mathematical criteria for truly random binary sequences have been intensively studied in the twentieth century see, e.g., review~\cite{Usp}. Unlike this, correlated binary sequences received much less attention. In particular, there is no known universal algorithm which may generate, like the  aforementioned inverse Fourier transformation, a binary sequence with arbitrary correlation function.

Several different methods
~\cite{CarStan,UYa,hod,nar1,nar2,muya,AllMemStepCor,Markov,genPre,gener}
are available  now for generation of binary sequences with limited
class of correlations. Each method  has its own advantages and
disadvantages. For example, the signum-generation method is
numerically simple and straightforward, but it was found in
Ref.~\cite{CarStan}, that the accuracy it reproduces the desired
correlation function is very limited. Much earlier the drawbacks of
this method were pointed out in Ref.~\cite{Watts}. Recently, the
signum-generation method was thoroughly examined, criticized and
improved in Ref.~\cite{gener}.

Some special classes of correlated functions can be reconstructed in
the binary sequences generated by linear transformation of a
\emph{binary additive Markov chain} using so-called {\emph{memory
function}}~\cite{UYa,muya,AllMemStepCor,Markov}. The memory function
shows the strength of the correlations and it is obtained from a
linear integral equation containing the pair correlator. Analytical
solution of this equation can be obtained in some special cases
only, which narrows the area of practical applicability of the
method.

Recently a new  \emph{filtering} method, which involves the
convolution procedure modified for generation of a binary sequence,
has been proposed~\cite{genPre}. The relation between the filtering
function, which serves as kernel of the convolution operator, and
the pair correlator turns out to be relatively simple. The advantage
of this method is that it requires less computation efforts to
generate a long sequence. Here we introduce a multi-step iterative
filtering method. After each step of modulation the generated
sequence becomes more correlated than the one obtained at the
previous step. We demonstrate that multiple iterations of initially
uncorrelated binary sequence may relatively quickly lead to a binary
sequence with desired pair correlator. The class of correlators
valid for this procedure includes not only exponentially decaying
correlators but also correlators with inverse-power-law decay. The
latter class of correlators are necessary for modeling DNA
sequences~{\cite{czir,CarStan,Chat,End,Bar,Bag}, anomalous
diffusion~\cite{Gri,RomSan}, and dynamics of complex
networks~\cite{Boc}.

The paper is organized as follows. In Section \ref{sec-FPM} we
discuss some general properties of binary sequences and reproduce
the key features of the filtering method. In Section
\ref{sec-TransCor}, we obtain a recursive relation between the
correlation functions after one step of iteration and display the
restrictions on the parameters, which provide the convergence of
iterative procedure. Here we also present a numerical example for
the direct problem -- generation of a binary sequence using a given
filtering function. In section \ref{Inverse} we consider the inverse
problem -- generation of a sequence with a given correlation
function, starting from an uncorrelated one. For the both problems
we demonstrate smooth gradual approach to the desired pair
correlator with  the number of consecutive iterations.

\section{Filtering Method}
\label{sec-FPM}

A random sequence $\{a\}$ of two symbols, 0 and 1,
\begin{equation}\label{BinWN}
a(n)=\{0,1\},\qquad n\in\textbf{\textbf{N}}=0,1,2,\ldots,
\end{equation}
can be characterized by the probability $p_1$ of occurring 1. Then, $p_0=1-p_1$ gives the probability of occurring 0. The probability $p_1$ coincides with the mean value
\begin{equation}\label{meanWN_a}
\overline{a}\equiv\overline{a(n)}=p_1{=\lim_{M\to\infty}\frac{1}{M}\sum_{n=0}^{M-1}a(n)}
\end{equation}
and averaging of any function $f(n)$ over the chain is defined as
\begin{equation}\label{meanWN_f}
\overline{f(a(n))}=f(0)p_0+f(1)p_1.
\end{equation}

We are interested in study of two-point correlation function $C_a(r)$
\begin{eqnarray}\label{CorWN}
&&C_a(r)=\overline{[a(n+r)-\overline{a}\,][a(n)-\overline{a}\,]}\nonumber\\[6pt]
&&=\lim_{M\to\infty}\frac{1}{M-r}\sum_{n=0}^{M-r-1}[a(n+r)-\bar{a}][a(n)-\bar{a}].
\end{eqnarray}

Our goal is to transform a random uncorrelated binary sequence $\{a\}$ into a binary correlated sequence $\{b\}$. To specify this transformation we introduce conditional probability $P(.|.)$ of occurring 1 at the $n$th place in a given sequence $\{b\}$. Then, the probability $P(.|.)$ is defined through the linear transformation
\begin{equation}\label{FPb}
P(b(n)=1|\{a\})=\overline{b}+\sum_{n'=-\infty}^{\infty}F(n-n')[a(n')-\overline{a}\,].
\end{equation}
Having the value of $P(.|.)$, the $n$th symbol is generated by drawing randomly a number from the interval [0,1]. If this number is smaller than $P(.|.)$, then $b(n)=1$, otherwise, $b(n)=0$. In fact, this procedure generates a statistical ensemble of the output sequences $\{b\}$ for each input sequence $\{a\}$. We assume that the input sequence $\{a\}$ is stationary. Since the linear transformation (\ref{FPb}) has a form of convolution, it generates the ensemble of output sequence $\{b\}$ which are also stationary. For the terms of the sequence $\{b\}$ any deterministic dependence of the probability $P(b(n)=1|\{a\})$ on its argument $n$ is suppressed. This property, which provides the stationarity of the generated sequence, means that the method of generation of the $n$th term is independent of $n$. Averaging of the probability $P(b(n)=1|\{a\})$ over statistical ensemble of sequences $\{a\}$ gives the mean value $\overline{b}$. Since $\overline{b}$ coincides with the mean probability for the symbol $1$ to appear in the whole sequence, the linear transformation (\ref{FPb}) preserves the ergodicity
\begin{equation}\label{Av-Alph}
\overline{b}=\overline{b(n)}=\overline{P(b(n)=1|\{a\})}.
\end{equation}

The \emph{filtering function} $F(n-n')$ in Eq.~(\ref{FPb}) describes the effect of the term $a(n')$ on the probability of $1$ occurring at the $n$th site in the sequence $\{b\}$. Positive values of $F(n)$ enhance existing fluctuations in the sequence $\{a\}$ and induce  persistent correlations in the sequence $\{b\}$, while negative values of the function $F(n')$ play the opposite smoothing role and induce anti-persistent correlations.

Being a probability, the function $P(b(n)=1|\{a\})$ takes the values between zero and one, $0\leqslant P(.|.)\leqslant1$. In terms of the filtering function this condition reads
\begin{equation}\label{FF-restr}
\sum_{n=-\infty}^{\infty}|F(n)|\leqslant\frac{\texttt{min}(\overline{b},1-\overline{b})}{\sqrt{\overline{b}(1-\overline{b})}}\,
\frac{\sqrt{\overline{a}(1-\overline{a})}}{\texttt{max}(\overline{a},1-\overline{a})}.
\end{equation}
For the unbiased sequence, ${\bar b}={\bar a}=1/2$ this inequality is reduced to
\begin{equation}\label{FF-weakrestr}
\sum_{n=-\infty}^{\infty}|F(n)|\leqslant1.
\end{equation}

Equation (\ref{FPb}) defines a set of probabilities which are
statistically independent. Therefore, the product
$P(b(n+r)=1|\{a\})P(b(n)=1|\{a\})$ gives the probability that 1
occurs at the $(n+r)$th and $n$th sites for a given realization of
the sequence $\{a\}$. For stationary sequence this product is
independent of $n$. Using the same arguments that were used to
obtain Eq.~(\ref{Av-Alph}), we get the following relation between
the binary correlation function and the on-site probabilities :
\begin{eqnarray}\label{Cb-Cfp}
&&C_b(r)\equiv\overline{[b(n+r)-\overline{b}\,][b(n)-\overline{b}\,]}\nonumber\\[6pt]
&&=\overline{[P(b(n+r)=1|\{a\})-\overline{b}\,][P(b(n)=1|\{a\})-\overline{b}\,]},
\nonumber\\
&&\qquad\mbox{for}\quad r\neq0.
\end{eqnarray}
This relation can be re-written through the binary correlation
functions of the input $a(n)$  and output $b(n)$  sequences. Using
Eqs.~(\ref{FPb}) and (\ref{Cb-Cfp}) and assuming that the mean
values and the variances of the two sequences $a(n)$ and $b(n)$ are
equal
\begin{equation}\label{a=b}
\overline{a}=\overline{b},\qquad\mbox{}\quad C_a(0)=C_b(0)=\overline{a}(1-\overline{a}),
\end{equation}
we obtain
\begin{equation}\label{CCF1}
K_{b}(r)=B_0\delta_{r,0}+\sum_{n,n'=-\infty}^{\infty}F(n)F(n')K_{a}(r+n-n').
\end{equation}
Here the normalized pair correlators $K_a(r)$ and $K_b(r)$ are defined as
\begin{equation}\label{Kcor-def}
K_{a,b}(r)=\frac{C_{a,b}(r)}{C_{a,b}(0)}
\end{equation}
and the constant $B_0$
\begin{equation}\label{B-def}
B_0 =1-\sum_{n,n'=-\infty}^{\infty}F(n)F(n')K_{a}(n-n')
\end{equation}
provides the normalization condition $K_{b}(0)=1$. It is worthwhile
mentioning that the fundamental relation (\ref{CCF1}) remains valid
even for the case when the input sequence $a(n)$ is correlated.
Indeed, so far we did not assume that $K_a(r)=\delta_{r,0}$.

The method of filtering probability, suggested here, is applicable
for two mutually inverse problems. One of them (the direct problem)
is numerical generation of a correlated sequence and calculation of
the correlation function $K_b(r)$ corresponding to a given filtering
function $F(n)$. Another one (the inverse problem), is
reconstruction of the filtering function $F(n)$ via a prescribed
correlator $K_b(r)$ of a random sequence. In the previous
publication~\cite{genPre} we analyzed Eq.~(\ref{CCF1}) when the
input random sequence was uncorrelated, i.e. $K_a(r)=\delta_{r,0}$.
It was shown that the ``discontinuity''\, of this equation at $r=0$
is the cause of some restrictions that strongly narrow the class of
correlation functions and corresponding random sequences which can
be considered by this method.

\section{Multi-Step Filtering}
\label{sec-TransCor}

In practice strongly correlated binary sequence can be hardly generated starting from a white-noise sequence. However, multiple application of the filtering transformation (\ref{FPb}) gradually increases the strength of correlations and may eventually lead to a binary sequence with desirable correlations. Therefore, it is worthwhile to consider a \emph{multi-step process of filtering transformations}. For each $(m+1)$th step the input sequence in Eq.~(\ref{FPb}) is the sequence obtained at the previous $m$th step with $K_{m}(r)$ being the correlator obtained after $m$ steps of filtering. The recurrence relation between the correlators $K_{m+1}(r)$ and $K_{m}(r)$ is readily obtained from Eqs.~(\ref{CCF1}), (\ref{B-def}),
\begin{equation}\label{CCFm}
K_{m+1}(r)=B_m\delta_{r,0}+\sum_{n,n'=-\infty}^{\infty}F(n)F(n')K_{m}(r+n-n'),
\end{equation}
\begin{equation}\label{Bm-def}
B_m=1-\sum_{n,n'=-\infty}^{\infty}F(n)F(n')K_{m}(n-n').
\end{equation}

Since we are dealing with stationary sequences, it is convenient to introduce the Fourier transform $\mathcal{K}(k)$ of the pair correlator $K(r)$, known as the randomness power spectrum,
\begin{eqnarray}\label{FT-K}
\mathcal{K}(k)=1+2\sum_{r=1}^{\infty}K(r)\cos(kr),\\[6pt]
K(r)=\frac{1}{\pi}\int_{0}^{\pi}\mathcal{K}(k)\cos(kr)dk.  \nonumber
\end{eqnarray}
The correlator $K(r)$ and its Fourier transform $\mathcal{K}(k)$ are both real and even functions of their arguments. Additionally, the power spectrum $\mathcal{K}(k)$ is a non-negative function of $k$ for any real random process. The recurrence relation (\ref{CCFm}) in the Fourier representation reads
\begin{equation}\label{Km(k)}
\mathcal{K}_{m+1}(k)=B_m+\mathcal{F}^2(k)\mathcal{K}_{m}(k),
\end{equation}
where $\mathcal{F}(k)$ is the Fourier transform of $F(n)$. If the recurrence relations (\ref{CCFm}) and (\ref{Km(k)}) converge in the limit $m\to\infty$ the following simple equation is obtained:
\begin{equation}\label{CF3}
\mathcal{K}(k)=B+\mathcal{F}^2(k)\mathcal{K}(k),
\end{equation}
where
\begin{eqnarray}\label{limK}
K(r)&=&\lim_{m\to\infty}{K}_{m}(r), \\
\mathcal{K}(k)&=&\lim_{m\to\infty}\mathcal{K}_{m}(k) \nonumber,\\
B&=&\lim _{m\to\infty}B_m.  \nonumber
\end{eqnarray}

It should be noted that for the class of correlators complying with the condition
\begin{equation}\label{conv_K_n}
\sum_{r=-\infty}^{\infty}|K(r)|<\infty,
\end{equation}
the existence of the limits (\ref{limK}) for the correlators is
guaranteed  by inequality (\ref{FF-weakrestr}). This follows
directly from the Banach fixed-point theorem provided that distance
between two correlation functions $K_{m+1}(r)$ and $K_{m}(r)$ is
defined as $\sum_{r}\big|K_{m+1}(r)-K_{m}(r)\big|$.

From Eq.~(\ref{CF3}) one obtains that a filtering function $F(n)$ gives rise to the correlator which cannot exceed the limiting value
\begin{equation}\label{SRCor-finFr}
K(r)=\frac{B}{2\pi}\int_{-\pi}^{\pi}\frac{\cos(kr)\,dk}{1-\mathcal{F}^2(k)}\,,
\end{equation}
where, in accordance with Eq.~(\ref{Kcor-def}), the normalization constant $B$ is
\begin{equation}\label{const-R-k}
B=\Big[\frac{1}{2\pi}\int_{-\pi}^\pi\frac{dk}{1-\mathcal{F}^2(k)}\Big]^{-1}.
\end{equation}
Here the correlator is completely defined by the filtering function $F(n)$. It is not the case for the correlator $\mathcal{K}_m(k)$ which, according to Eq.~(\ref{CCFm}) depends also on the correlator at the preceding step and, thus, on the correlator of the initial input sequence. However, after infinite number of mappings given by Eq.~(\ref{CCFm}) the output sequence riches the "limit cycle" where the memory on "initial condition" is
completely lost.

\protect\begin{figure}
\begin{center}
\includegraphics[width=8cm]{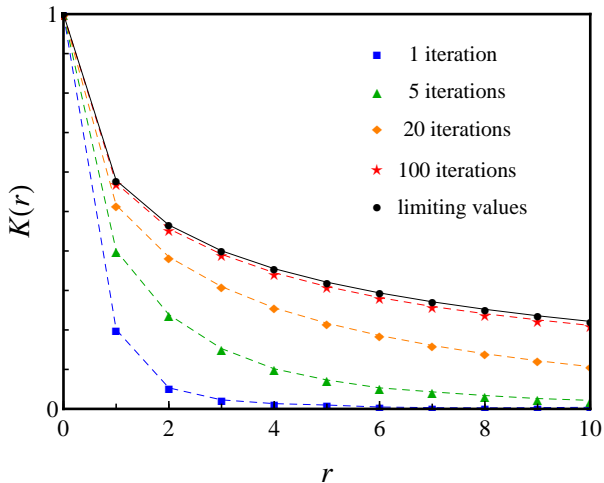}
\caption{\label{Ex-Fig} (Color online) Correlation functions of the binary sequences generated using the filtering function (\ref{Ex-1}) after $m=1,5,20$, and 100 iterations. The strongest correlations are observed for the solid black curve ($m=\infty$) corresponding to Eq.~(\ref{SRCor-finFr}). The length of the sequence is $10^5$ and the parameter $\alpha=0.2$.}
\end{center}
\end{figure}

In order to illustrate the effectiveness of the proposed iterative method we perform numerical simulations and generate a set of  correlated sequences, using the filtering function with a power-law decay
\begin{eqnarray}\label{Ex-1}
&&F(n)=\sqrt{\frac{{\alpha}}{2}}\,\frac{1-(-1)^n}{\pi n^2}\,,\qquad F(0)=\frac{\pi}{2}\sqrt{\frac{{\alpha}}{2}}\,,\nonumber\\[6pt]
&&\mathcal{F}(k)=\sqrt{\frac{{\alpha}}{2}}\big(\pi-|k|\big)\,,\qquad|k|\leqslant\pi.
\end{eqnarray}
Here, due to the requirement (\ref{FF-weakrestr}), the constant $\alpha$ lies within the interval
\begin{equation}\label{a-sq}
0<{\alpha<\alpha_{\rm max}}=\frac{2}{\pi^2}\approx 0.203.
\end{equation}
The starting binary input sequence $\{a\}$ is delta-correlated white-noise
\begin{eqnarray}\label{K-WN}
&&K_a(r)\equiv K_{0}(r)=\delta_{r,0}\,,\nonumber\\[6pt]
&&\mathcal{K}_a(k)\equiv \mathcal{K}_{0}(k)=1.
\end{eqnarray}

The correlators of the sequences generated after 1, 20 and 100 iterations and the correlator (\ref{SRCor-finFr}) of the sequence corresponding to infinite iterations are shown on Fig.~\ref{Ex-Fig}. The lowest curve presents the correlator of the sequence $\{b\}$ obtained after the first filtering of the uncorrelated sequence $\{a\}$. This result is in complete correspondence with the analytical expression
\begin{equation}\label{Filter_power corr}
K_{b}(r)\equiv K_{1}(r)=\delta_{r,0}+\frac{{\alpha}}{r^2} (1-\delta_{r,0})
\end{equation}
that directly follows from Eqs.~(\ref{CCF1}), (\ref{B-def}) and (\ref{Ex-1}). One can see that single-step filtering of the white-noise with the filtering function (\ref{Ex-1}) generates a sequence with the power-law decaying correlations. A set of the curves in Fig.~\ref{Ex-Fig} shows  increase of the strength of correlations with the number of iterations and gradual approach to the limiting value $K(r)$ defined by Eq.~(\ref{SRCor-finFr}).

Gradual increase of strength of correlations with the number of iterations $m$ can be estimated for $r \gg 1$. In this limit Eq.~(\ref{SRCor-finFr}) exhibits $1/r^2$ behavior and the rate of the correlations increase can be characterized by the ratio
\begin{equation}\label{K-assymp}
\frac{K(r)}{K_{1}(r)}\approx\frac{4\sqrt{2\alpha}}{\pi^3 (\alpha_{max}-\alpha)^{2}} \left(\ln\frac{\alpha_{max}+\alpha}{{\alpha_{max}-\alpha}}\right)^{-1},\quad|r|\gg1.
\end{equation}
This ratio exceeds $1$ for all the values of $\alpha$ from the interval (\ref{a-sq}). For $\alpha$ close to $\alpha_{max}$ it diverges, showing infinite rate. Thus, each iteration of the transformation (\ref{CCFm}) indeed increases the strength of correlations in the generated binary sequence. These iterations, however, do not affect the rate the correlations decay with the distance $r$, which remains the same, $K_{m}(r)\sim1/r^{2}$ at $r\gg1$ for any $m$.

\section{Inverse Problem}
\label{Inverse}

In this Section we study the inverse problem which is formulated as generation of random binary sequence with prescribed correlation function, starting from delta-correlated white noise Eqs.~(\ref{K-WN}).

From Eq.~(\ref{CF3}) one can readily obtain the filtering function,
\begin{equation}\label{FB(n)}
F_B(n)=\frac{1}{2\pi}\int_{-\pi}^{\pi}dk\,\cos(kn)\sqrt{1-\frac{B}{\mathcal{K}(k)}}\,,
\end{equation}
which generates a sequence if the binary correlator $K(r)$ and its Fourier transform $\mathcal{K}(k)$ are known. Here the constant $B$ should be considered as a free parameter, unlike Eq.~(\ref{SRCor-finFr}) where it is calculated from the normalization condition (\ref{const-R-k}). Since the filtering function $F_B(n)$ takes only the real values, the parameter $B$ cannot exceed $\mathcal{K}(k)$. The lower bound for $B$ is obtained from the following chain of relations, including Eq.~(\ref{FF-weakrestr}):
\begin{eqnarray}
&&\sqrt{1-\frac{B}{\mathcal{K}(k)}}=\mathcal{F}_B(k)=\sum_{n=-\infty}^{\infty}F_B(n)\exp(-ikn)\nonumber\\[6pt]
&&\leqslant\sum_{n=-\infty}^\infty|F_B(n)|\leqslant1.
\end{eqnarray}
Comparing the first and last terms in this chain we conclude that
the parameter $B$ should be positive, i.e., it is chosen from the
interval
\begin{equation}\label{BRest}
0< B<\mathcal{K}(k),\qquad|k|\leqslant\pi.
\end{equation}
We exclude $B=0$ from the allowed values since it is obvious that the corresponding filtering function $F_0(n)=\delta_{n,0}$ describes the identity transformation. Accordingly, a transformation with $B\to0$ turns out to be  a near-identity transformation.

Formally any choice of $B$ from the interval (\ref{BRest}) leads to a filtering function $F_B(n)$ which generates (after infinite number of iterations) a binary sequence with the same correlation function $K(r)$. However, it will be shown below that the proper choice of $B$ may optimize the convergence of the series of the intermediate correlators $K_{m}(r)$ to the desired  $K(r)$ and rich better accuracy for the same number of iterations in the procedure.

\subsection{Near-identity transformations}

In this Subsection we study the filtering procedure with sufficiently small $B$. More specifically, we assume
\begin{equation}
0<B\ll\min_{|k|\leqslant\pi}\mathcal{K}(k).
\end{equation}
Expanding Eq.~(\ref{FB(n)}) over $B$ and keeping the linear term we
obtain the filtering function for near-identity transformation
\begin{eqnarray}\label{f-expr}
F_B(n)&=&\delta_{n,0}-Bf(n),\nonumber\\[6pt]
f(n)&=&\frac{1}{4\pi}\int_{-\pi}^{\pi}\frac{\cos(kn)}{\mathcal{K}(k)}\,dk.
\end{eqnarray}
In terms of $f(n)$ the condition (\ref{FF-weakrestr}) reduces to
\begin{equation}\label{smallB-FF}
\frac{1}{2}\,f(0)-\sum_{n=1}^\infty|f(n)|\geqslant0.
\end{equation}
This inequality is used for optimization of the numerical procedure.
It should be reminded that inequality (\ref{smallB-FF}) coincides
with restriction (\ref{FF-weakrestr}) for small $B$ only, and it is
weaker then Eq.~(\ref{FF-weakrestr}) otherwise.

\subsection{Exponentially decaying correlation}

Short-range correlations can be approximated by exponentially decreasing correlator,
\begin{eqnarray}
&&K(r)=\exp(-\gamma|r|)\,,\qquad\gamma>0\,,\label{CorrExp}\\[6pt]
&&\mathcal{K}(k)=\frac{\sinh\gamma}{\cosh\gamma-\cos k}\,,\label{FCorrExp}
\end{eqnarray}
where $1/\gamma$ is the correlation radius. Since $\mathcal{K}(k)$ reaches its minimum at $k=\pi$, inequality (\ref{BRest}) gets the following form:
\begin{equation}
0<B<\tanh\dfrac{\gamma}{2}.
\end{equation}
For small $B\ll\tanh(\gamma/2)$ the near-identity transformation defined by Eq.~(\ref{f-expr}) yields
\begin{equation}
f(n)=\dfrac{1}{4\sinh\gamma}(\delta_{n,1}-2\delta_{n,0}\cosh\gamma+\delta_{n,-1}).
\end{equation}
Then condition (\ref{smallB-FF}) becomes $\tanh(\gamma/2)>0$, i.e.
the filtration procedure is allowed for any small $B$. Application
of the multi-step method allows generation of binary sequences with
correlator (\ref{CorrExp}) for \emph{any} value of  $\gamma>0$.
Unlike this, the single-step filtering \cite{genPre} is not
applicable for $\gamma<\gamma_{cr}\approx1.60$.

The correlated binary sequence of length $10^6$ was generated using
the iterative filtering procedure with $F_B(n)$ given by
Eq.~(\ref{FB(n)}) for two different values of $B$ and $\gamma=0.5$.
The corresponding correlators for $B=0.1$ and $B=0.02$ are shown in
Fig.~\ref{ExpFig}. For both values of $B$ the correlators in the
generated sequences gradually approach (with the number of
iterations) the exponential correlator (\ref{CorrExp}). However, the
rate of convergence depends on $B$. As one may expect, smaller
values of $B$ leads to slower convergence.

\protect\begin{figure}
\begin{center}
\includegraphics[width=8cm]{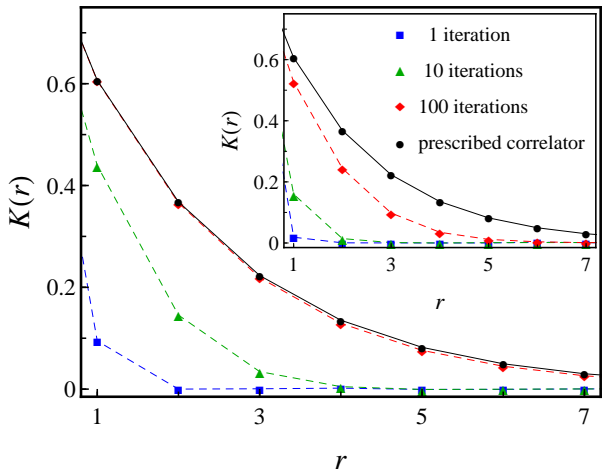}
\caption{\label{ExpFig}(Color online) The correlator of the binary sequence generated with $B=0.1$ (main panel) and $B=0.02$ (insert) after $1$, $10$, and $100$ filtering procedures. For $B=0.1$ the correlator obtained after 100 iterations practically coincides with the prescribed correlator (\ref{CorrExp}). The length of the sequence is $10^6$, and the parameter $\gamma=0.5$.}
\end{center}
\end{figure}

\subsection{Power-law correlator}

A complex system or a system close to its critical state is
characterized by long-range correlations, i.e., the correlator decays
as inverse power law
\begin{eqnarray}
&&\mkern -40mu K(r)=\delta_{r,0}+\frac{{\alpha}}{|r|^p}(1-\delta_{r,0})\,,\qquad p>1,\label{CorrPow}\\[6pt]
&&\mkern -40mu \mathcal{K}(k)=1+2\alpha Z_p(k)\,,\qquad Z_p(k)=\sum_{r=1}^\infty\frac{\cos(kr)}{r^p}\,.\label{FCorrPow}
\end{eqnarray}
The constant ${\alpha}=K(1)$ is chosen to ensure that the condition $\mathcal{K}(k)\geqslant0$ is satisfied within the whole interval $|k|\leqslant\pi$. Taking into account that $Z_p(0)=\zeta(p)$,  where $\zeta(p)$ is the Riemann zeta function, the following interval for $\alpha$ is obtained:
\begin{equation}\label{a-pow-cond}
-\frac{1}{2\zeta(p)}<\alpha<\frac{1}{2(1-2^{1-p})\zeta(p)}.
\end{equation}

The filtering function (\ref{FB(n)}) corresponding to the power-low correlator (\ref{CorrPow}) reads
\begin{equation}\label{FB(n)-power}
F_B(n)=\frac{1}{2\pi}\int_{-\pi}^{\pi}dk\,\cos(kn)\sqrt{1-\frac{B}{1+2\alpha Z_p(k)}}\,.
\end{equation}
Since the function $F_B(n)$ must satisfy inequality
(\ref{FF-weakrestr}), there is one more restriction on the parameter
$\alpha$. It is easier to obtain this restriction using
Eq.~(\ref{smallB-FF}) which defines the filtering function for small
$B$,
\begin{equation}
0<B\ll \min\mathcal{K}(k)=1-2\alpha(1-2^{1-p})\zeta(p),
\end{equation}
rather than the general condition (\ref{FF-weakrestr}).
Equation~(\ref{f-expr}) gives
\begin{equation}
f(n)=\frac{1}{4\pi}\int_{-\pi}^{\pi}\frac{\cos(kn)\,dk}{1+2\alpha Z_p(k)}\,.
\end{equation}
Substituting this result to inequality (\ref{smallB-FF}) we obtain
numerically the maximal value of~$\alpha$ allowed for each value
of~$p$. This dependence is plotted in the inset of
Fig.~\ref{PowFig}.

\begin{figure}
\begin{center}
\includegraphics[width=8cm]{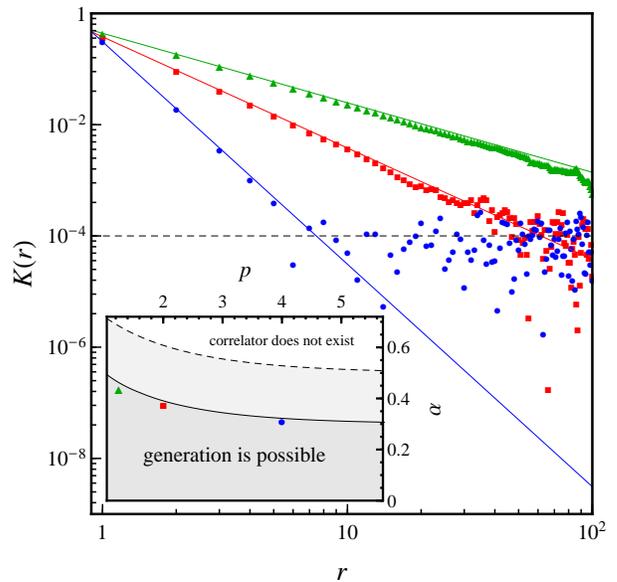}
\caption{\label{PowFig}(Color online) Numerically calculated correlation function $K(r)$ of the sequences generated by the filtering function (\ref{FB(n)-power}) for $p_1=1.2$, $\alpha_1=0.44$ (triangles), $p_2=2$, $\alpha_2=0.38$ (squares), and $p_3=4$, $\alpha_3=0.31$ (circles) shown in logarithmic scale. Solid straight lines represent the prescribed correlators~(\ref{CorrPow}). The number of filtering iterations is $m=200$, the sequence length is $M=10^{8}$ elements, and $B=0.05$. The dashed horizontal line at $K(r)=10^{-4}$ indicate the fluctuation border $1/\sqrt{M}$ of the correlator. The inset shows the possible values of the parameter $\alpha$. Solid line is the upper limit for $\alpha$ imposed by the condition (\ref{smallB-FF}). For $\alpha$'s below this limit the multi-step filtering procedure converges. The dashed line is obtained from the right condition (\ref{a-pow-cond}). Above this line the function $K(r)$ given by Eq.~(\ref{FCorrPow}) cannot serve as correlator of a random sequence. The points marked by triangle, square, and circle in the inset represent the particular values of $p$ and $\alpha$ chosen for generation of the data shown in the main panel. The lower limit for $\alpha$ given by Eq.~(\ref{a-pow-cond}) lies in the region of negative values and it is not shown here.}
\end{center}
\end{figure}

While the upper limit for $\alpha$ was obtained assuming that $B\ll1$, it remains valid for any $B$. This conclusion follows from our numerical simulations which show that smaller $B$'s always allow larger values of $\alpha$.

We apply the iterative method for generation of three correlated sequences with correlator (\ref{CorrPow}) and parameters $p_1=1.2$, $\alpha_1=0.45$; $p_2=2$, $\alpha_2=0.38$; and $p_3=4$, $\alpha_3=0.3$. The values of the parameter $\alpha$ are specially chosen to be close to the maximum values allowed by Eq.~(\ref{smallB-FF}): $\alpha_{1max}\approx0.461$, $\alpha_{2max}\approx0.389$ and $\alpha_{3max}\approx0.322$.  Each of the generated sequences contains $10^8$ terms which were obtained after $m =200$ steps of filtering iterations. The correlators of the generated sequences were calculated numerically. The results  are shown in Fig.~\ref{PowFig}. It is clear that the correlators of the generated sequences follow well the prescribed power-law decay up to $K(r) \approx 10^{-4}$. When the correlations fall below this critical value, the numerical results exhibit strong fluctuations and the agreement with the power-law decay is lost. This discrepancy is due to the finite length of the generated sequences. It follows from the law of large numbers that correlation function, as any other averaged statistical characteristic of a random sequence, is a deterministic
quantity only if the sequence length is infinite. For a finite length $M$ the correlation function becomes a $M$-dependent fluctuating quantity. The amplitude of the fluctuations decays with $M$ as $1/\sqrt{M}$. For the sequences with length $M=10^8$ the fluctuations are $\sim 10^{-4}$, i.e. they become essential  when the correlator drops below $K(r) = 10^{-4}$. As we can see from Fig.~\ref{PowFig}, this value is, indeed, establishes the precision limit for numerical reproduction of the correlator $K(r)$.

\subsection{Correlator with colored-noise spectral density
\label{color}}

There is a special class of correlators $K(r)$ which decay very slowly with $|r|$. Here we consider binary sequences with colored-noise spectrum
\begin{equation}\label{Kor}
\mathcal{K}(k)=(1-\beta)\left(\frac{\pi}{|k|}\right)^\beta,\qquad0<\beta\leqslant1.
\end{equation}
The factor $1-\beta$ is introduced for normalization $K(r=0)=1$ of the corresponding correlator in real space, where it decays as inverse power-law for $0<\beta\leqslant1$,
\begin{equation}
K(r)\approx\big(\pi |r|\big)^{\beta -1}\Gamma(2-\beta )\sin\frac{\pi\beta}{2},\quad|r|\to\infty,
\end{equation}
It, however, decays exponentially for $\beta>1$ \cite{Garcia}, after appropriate cut-off at $k \rightarrow 0$.

Since the correlator $K(r)$ decays slowly, one might expect the filtering function
\begin{equation}\label{Fmem}
F_B(n)=\int_{0}^{1}\cos(\pi x n)\sqrt{1-\frac{B}{1-\beta}{x}^\beta}\,dx\,.
\end{equation}
to decrease very slowly, as well.
However, due to oscillations of the integrand, the asymptotics of $F_B(n)$ falls off rapidly enough at $|n|\to\infty$,
\begin{eqnarray}
\label{asymp}
&& F_B(n)\approx \frac{B \beta}{2 \sqrt{1-\beta}|\pi n|^{1+\beta}}\left[\frac{\Gamma(\beta) \sin(\pi \beta/2)}{\sqrt{1-\beta}} \right. \nonumber \\
&& - \left. \frac{(-1)^n}
{\sqrt{1-\beta-B} |\pi n|^{1-\beta}} \right],
\end{eqnarray}
providing convergence of the series in Eq.~(\ref{FF-weakrestr}). Two terms in this asymptotics originate from the contributions of the end points, $x=0$ and $x=1$. Usually the point $x=0$ gives the principal contribution described by the first term. The second term decays faster with $n$ and we show below that it may play a role only in the case of some special relation between $B$ and $\beta$.

Convergence of the series $\sum_n |F_B(n)|$ is not sufficient for
the convergence of the multi-step filtering procedure. The latter is
guaranteed by the condition (\ref{FF-weakrestr}), i.e. this series
cannot exceed one. To calculate the series we assume that for any
integer $n$ the value of the filtering function is positive,
$F_B(n)>0$. This is true for $|n| \rightarrow \infty$, as it can be
seen from asymptotics (\ref{asymp}). For moderate $n$ the value of
the integral is dominated by positive contribution from the interval
$0 \leq x \leq 1/2n$. The rest of the region of integration usually
gives smaller (by its absolute value) contribution because of the
decreasing square-root factor and oscillations. Exceptions from this
rule are considered below. For positive values of $F_B(n)$ the sign
of absolute value can be omitted in Eq.~(\ref{FF-weakrestr}) and the
sum over $n$ can be calculated using the identity
\begin{equation}\label{Poisson}
\sum_{n=-\infty}^{\infty}\cos(\pi nx)=2\delta(x),\qquad|x|<1.
\end{equation}
Now inequality (\ref{FF-weakrestr}) reads
\begin{eqnarray}
\label{sum}
&&\sum_{n=- \infty}^{\infty} F_B(n)= \int_0^1 dx \sqrt {1-\frac{B}{1-\beta}x^{\beta}} \sum_{n=- \infty}^{\infty} \cos(\pi n x)
 \nonumber \\
&&= 2 \int_0^1 \delta(x) \sqrt {1-\frac{B}{1-\beta}x^{\beta}} dx = 1.
\end{eqnarray}
Thus, inequality (\ref{FF-weakrestr}) still remains true, for any
$\beta$ and $B$ satisfying the condition $B/(1-\beta) \leqslant1$,
that ensure convergence of the multi-step procedure.

\begin{figure}
\begin{center}
\includegraphics[width=8cm]{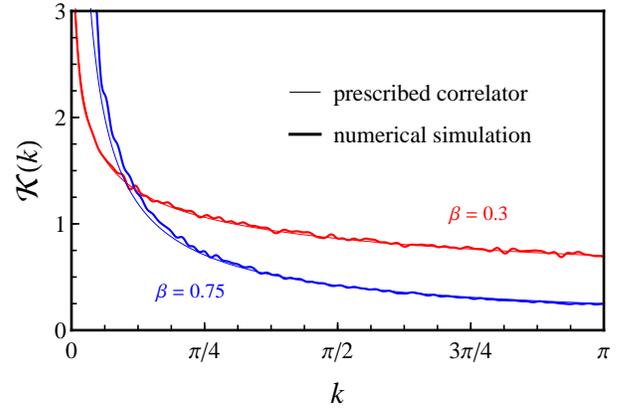}
\caption{\label{spectrum}(Color online) The power spectra $\mathcal{K}(k)$
of the binary sequences generated  using the filtering function (\ref{Fmem}) with $\beta=0.3$ ($10$ iterations with $B=0.6$) and $\beta=0.75$ ($20,000$ iterations with $B=0.13$). Thin lines are for the prescribed spectrum (\ref{Kor}). Thick lines are the results of the numerically calculated power spectra of the generated random sequences of the length $\sim 10^{6}$ terms.}
\end{center}
\end{figure}

In numerical simulations larger values of $B$ are preferential since
they provide faster convergence. Therefore, in our numerical
calculations we choose the value of $B$ which is a bit less  than
that allowed by inequality $B/(1-\beta) \leqslant1$.
Figure~\ref{spectrum} shows two power spectra for of the binary
sequences generated with $\beta=0.3$, $B=0.6$ and $\beta=0.75$,
$B=0.13$. While both spectra fit equally well the prescribed
analytical curves, the numerical efforts they required are quite
different. The number of iterations to generate a sequence with
$\beta = 0.3$ was only 10. Unlike this, to generate a correlated
sequence with $\beta=0.75$ we performed $2\cdot10^{4}$ steps to rich
approximately the same accuracy. This drastic difference is due to
the smaller values of $B$ which are allowed when $\beta$ becomes
relatively close to one.

It is important to note that although the larger values of the parameter $B$  provide faster convergence, they cannot be taken very close to the limiting value $1- \beta$. It is hard to evaluate  the width of this prohibited interval in general case but the fact that such interval does exist can be seen from Eq.~(\ref{asymp}). For even $n$'s the second term there gives negative contribution, which may overcome the positive first term, if $\sqrt{1-\beta-B }\rightarrow 0$. Presence of even one negative term in the series (\ref{sum}) means that the original series with absolute values in  Eq.~(\ref{FF-weakrestr}) exceeds one, i.e. the multi-step filtering procedure diverges.

For values of $B$ close to $1-\beta$ asymptotics (\ref{asymp})
becomes invalid and it must be replaced by
\begin{eqnarray}
\label{asymp2}
&&F_{B=1-\beta}(n) = \int_0^1 \cos(\pi x n) \sqrt{1-x^{\beta}}\, dx  \nonumber \\
&& \approx \frac{\sin(\pi \beta /2) \Gamma(1+ \beta)}{2|\pi n|^{1+\beta}} - \frac{(-1)^n \sqrt{\beta/2}}{2 \pi |n|^{3/2}}.
\end{eqnarray}
The width of the prohibited interval turns out to be narrow for
$\beta$ not so close to $1$ and its presence does not impose
practical limitation on the proposed method of generation of binary
colored noise. For example, for $\beta = 0.3$ the width of the
prohibited interval is exactly zero. This can be seen by evaluating
the function $F_{1-\beta}(n)$ numerically or using asymptotics
(\ref{asymp2}). It remains positive for any integer $n$ and $\beta =
0.3$. However, for $\beta = 0.75$ the function $F_{1-\beta}(n)$
takes negative values for any even $n \geqslant 2$. Numerical
evaluation of $F_B(n)$ shows that this function is positive for $B
\leqslant 0.18$, i.e. the width of the prohibited interval is
0.25-0.18 = 0.07.

It is worth mentioning that the discrete Fourier transform algorithm
for generation of a \emph{non-binary} sequence with spectral density
(\ref{Kor}) for any value of $\beta$ is widely used in studies of
fractional Brownian motion \cite{Weiss}. It was recently pointed out
that for $\beta >1$ this algorithm generates sequences that are not
truly random \cite{Nancy}. It turns out that in the thermodynamic
limit the correlator $K(r)$ does not vanish at $r \rightarrow
\infty$ approaching a finite negative value. Domination of these
anticorrelations is a source of extended quantum states predicted
for the tight-binding model with diagonal disorder \cite{Lyra}.

\section{Conclusion}

We propose a new multi-step iterative method for generation of correlated binary sequences  with a prescribed pair correlation function. The method is based on the multiple filtering procedure when each next step of filtering generates a sequence with stronger correlations than those generated at the previous steps. We demonstrate the applicability of the new method by generating long binary sequences with exponential, inverse power-law decaying correlators, and binary sequences with colored noise. The latter case is a challenging problem since the power spectrum density $\mathcal{K}(k)$ is represented by a slowly decaying function, $\mathcal{K}(k)\sim|k|^{-\beta}$ with $0<\beta\leqslant 1$. We are not aware of any other method which generates a binary colored-noise sequence.

\begin{acknowledgments}
We acknowledge support form the SEP-CONACYT (M\'exico) under grant No. CB-2011-01-166382.
\end{acknowledgments}



\end{document}